# LHC AND VLHC BASED ep COLLIDERS: e-LINAC vs e-RING

L. Gladilin[1], H. Karadeniz[2,3], E. Recepoglu[2,*], S. Sultansoy[3,4]

[1] Skobeltsyn Institute of Nuclear Physics, Moscow State University, Moscow, Russia.

[2] Ankara Nuclear Research and Training Center, 06100 Besevler, Ankara, Turkey.

[3] Dept. of Physics, Faculty of Arts and Sciences, Gazi University, 06500 Teknikokullar, Ankara, Turkey.

[4] Institute of Physics, Academy of Sciences, H. Cavid Ave. 33, Baku, Azerbaijan.

**Abstract**

Linac-ring analogues of the LHC and VLHC based standard type ep collider proposals are discussed. It is shown that sufficiently high luminosities can be obtained with TESLA-like linacs, whereas essential modifications are required for CLIC technology. The physics search potential of proposed ep colliders is demonstrated using pair production of heavy quarks as an example.

**Keywords:** Lepton, hadron, luminosity, ep collider.

## 1. Introduction

It is known that lepton-hadron collisions have been playing a crucial role in exploration of deep inside of matter. For example, the quark-parton model was originated from investigation of electron-nucleon scattering. The HERA with $\sqrt{s_{ep}} \approx 0.3$ TeV has opened a new era in this field extending the kinematics region by two orders both in high $Q^2$ and small x with respect to fixed target experiments. However, the region of sufficiently small x ($\leq 10^{-5}$) and simultaneously high $Q^2$ (= 10 GeV²), where saturation of parton densities should manifest itself, is not currently achievable. The investigation of physics phenomena at extreme small x but sufficiently high $Q^2$ is very important for understanding the nature of strong interactions at all levels from nucleus to partons. At the same time, the results from lepton-hadron colliders are necessary for adequate interpretation of physics at future hadron colliders. Concerning LHC, which hopefully will start in 2007, an $\sqrt{s} \approx 1\,\text{TeV}$ ep collider will be very useful in earlier 2010's when precision era at LHC will begin.

---

[*] Corresponding author: recepoglu@taek.gov.tr

Today, linac-ring type machines seem to be the main way to TeV scale in lepton-hadron collisions (see [1] and references therein). Construction of future linear collider or a special e-linac tangentially to existing (HERA, TEVATRON, RHIC) or planned (LHC, VLHC) hadron rings will provide a number of new powerful tools in addition to ep and eA options:

- TeV scale γp [2] and γA [3] colliders. In this case high energy electron beam will be converted into photon beam using Compton back scattering of laser photons on ultra-relativistic electrons [4]. It should be noted that photon-hadron options can not be realized on the base of standard (ring-ring) type electron-hadron colliders. (see arguments given in [2])
- FEL-Nucleus colliders [5]. In this case (a part of) e-linac will be used for production of keV energy laser beam. Let us mentioned that FEL-Nucleus colliders satisfy all requirements on ideal photon source for nuclear resonance fluorescence experiments [6].

On the other hand, there are several standard (ring-ring type) ep collider proposals with $\sqrt{s_{ep}} > 1\,\text{TeV}$. The first one is an ep option for LHC [7]. This proposal, which assumes a construction of 67.3 GeV electron ring in the LHC tunnel, is considered as a part of the LHC programme in [8]. Concerning the VLHC based ep collider, a construction of 180 GeV e-ring in the VLHC tunnel is proposed in [9].

In this paper we consider linac-ring analogs of the LHC and VLHC based standard type ep colliders mentioned above. Two basic assumptions are made:

1- Linac and ring beams have the same electron energy,
2- Linac beam power is equal to e-ring synchrotron radiation power.

Main limitations on linac-ring type ep collider parameters are discussed in section 2. Comparison of the LHC and VLHC based linac-ring type and standard type ep colliders is performed in sections 3 and 4, respectively. As an example of physics search potential we consider pair production of heavy quarks ($\bar{c}c$ and $\bar{b}b$) in section 5. Finally, we give some concluding remarks in section 6.

## 2. General Consideration

There are two most important collider parameters from physicist point of view, namely, center of mass energy and luminosity. The center of mass energy determines the scale of resolved dimension and achievable masses of new particles. The luminosity multiplying by corresponding cross-section determines number of available events. In addition, beam polarization, energy spread, collision frequency and luminosity per collision could be important for different phenomena.

The center of mass energy is given by $\sqrt{s} = 2\sqrt{E_e E_p}$ for ultra-relativistic head-on colliding particles. The most transparent expression for the luminosity of linac-ring type ep colliders is [10]:

$$L_{ep} = \frac{1}{4\pi} \frac{P_e}{E_e} \frac{N_p}{\varepsilon_p^N} \frac{\gamma_p}{\beta_p^*} \qquad (1)$$

for round, transversely matched beams with the same bunch spacing. Here, $E_e$ is the energy of electrons, $P_e$ is electron beam power, $N_p$ and $\varepsilon_p^N$ are the number of particles in proton bunch and normalized emittance of proton beam, $\gamma_p$ is the Lorentz factor and $\beta_p^*$ is amplitude function of proton beam at interaction point.

The first restrictive limitation for electron beam is beam power

$$P_e = N_e E_e n_b f_{rep} \qquad (2)$$

where $N_e$ is the number of particles in electron bunch, $n_b$ is the number of bunches in linac pulse and $f_{rep}$ is repetition rate of the linac. Taking into account the acceleration efficiency, reasonable value of $P_e$ is several tens MW.

The maximum number of electrons per bunch is determined by the beam-beam tune shift limit of the proton beam

$$\Delta Q_p = \frac{N_e r_0 \beta_p^*}{2\pi\gamma_p \sigma_{xe}(\sigma_{xe}+\sigma_{ye})} \quad (3)$$

where $r_0 = 1.54 \cdot 10^{-18}$ m is the classical radius of proton. $\sigma_{xe}$ and $\sigma_{ye}$ are horizontal and vertical sizes of electron beam at interaction point. Generally accepted beam-beam tune shift value for protons in the case of ring-ring colliders is ? $Q_p$ = 0.003. This limit value can be a little bit larger for linac-ring type colliders.

Disruption parameter for electrons is given by

$$D = \frac{2N_p r_e}{\gamma_e} \frac{\sigma_{zp}}{\sigma_{xp}(\sigma_{xp}+\sigma_{yp})} \quad (4)$$

where $r_e$ is the classical electron radius, $\sigma_{zp}$ is proton bunch length, $\sigma_{xp}$ and $\sigma_{xy}$ are the horizontal and vertical proton beam sizes.

The most important limitation on proton beam comes from intrabeam scattering (IBS), which leads to emittance growth. We assume that IBS growth time $\tau_{IBS} \geq 1$ hour is acceptable for linac-ring type colliders. For comparison, filling time is 7.5 min. and acceleration period is 1200 s for the LHC proton beam. In our calculations we use formulas from [11].

**3. The LHC Based ep Collider (QCD Explorer)**

As mentioned above, the standard ep option for LHC [7] assumes a construction of 67.3 GeV e-ring in the LHC tunnel and it is considered as a part of the LHC programme [8]. Parameters of electron and proton beams for this option are given in Table 1. It is seen that center of mass energy $\sqrt{s_{ep}}$ = 1.37 TeV and luminosity $L_{ep} = 1.2 \cdot 10^{32}$ cm$^{-2}$s$^{-1}$ will be achieved.

| Electron beam parameters | |
|---|---|
| Energy $E_e$ (GeV) | 67.3 |
| Bunch Population, $N_e$ | $6.4 \cdot 10^{10}$ |
| Emittance, $\varepsilon_e$ (nm) | 9.5/2.9 |
| Beta functions, $\beta_{xe}/\beta_{ye}$, (m) | 0.85/0.26 |
| Beam-beam tune shifts, $\xi_x/\xi_y$ | 0.027/0.027 |
| Radiation power W [MW] | 34.5 |
| Proton beam parameter | |
| Energy $E_p$ (TeV) | 7 |
| Bunch Population, $N_p$ | $10^{11}$ |
| Emittance, $\varepsilon_p$ (nm) | 0.5 |
| Beta functions, $\beta_{xp}/\beta_{yp}$, (m) | 16/1.50 |
| Beam-beam tune shifts, $\xi_x/\xi_y$ | 0.0032/0.0010 |
| Collider parameters | |
| Center of mass energy, $\sqrt{s}$ (TeV) | 1.37 |
| Luminosity ($10^{32}$ cm$^{-2}$s$^{-1}$) | 1.2 |

Table 1. Parameters for standard type LHC ep collider [6].

Let us consider the use of e-linac instead of e-ring with ~27 km circumference. With $E_e = 67.3$ GeV and $P_e = 34.5$ MW and nominal LHC proton beam parameters ($N_p = 1.1 \cdot 10^{11}$, $\varepsilon_p = 0.5$ nm, $\beta_p^* = 0.5$ m [12]) we obtain for linac-ring option $L_{ep} = 1.1 \cdot 10^{31}$ cm$^{-2}$s$^{-1}$ according to Eq. (1). If one choose the THERA proton beam parameters [13], namely, $N_p = 10^{11}$, $\varepsilon_p^N = 10^{-6}$ m and $\beta_p^* = 10$ cm the luminosity for "ideal" e-linac becomes $L_{ep} = 1.9 \cdot 10^{32}$ cm$^{-2}$s$^{-1}$.

Concerning the "real" e-linac technologies we consider TESLA and CLIC proposals. Parameters of the TESLA (THERA option [13]) and CLIC [14] e-beams are given in Table II. It is seen from Table III that in the TESLA case one can use all e-bunches, whereas only ~3% of the CLIC e-bunches will collide with the LHC proton bunches. (Let us mentioned that superbunch option for the LHC could give opportunity to utilize all CLIC bunches [15] but this opportunity requires a radical modification of whole LHC stages from injector to main ring). With nominal LHC parameters we obtain $L_{ep} = 1.9 \cdot 10^{30}$ cm$^{-2}$s$^{-1}$ for "TESLA" and $L_{ep} = 1.4 \cdot 10^{28}$ cm$^{-2}$s$^{-1}$ for "CLIC" (see Table III). With THERA like modification of the LHC proton beam, the luminosity values become $L_{ep} = 3.3 \cdot 10^{31}$ cm$^{-2}$s$^{-1}$ and $L_{ep} = 2.3 \cdot 10^{29} \, cm^{-2} s^{-1}$ respectively (see Table IV). It is seen that a factor of ~3.5 for TESLA technology and a factor of ~ 500 for CLIC technology are needed in order to achieve a luminosity $L_{ep} = 1.2 \cdot 10^{32} \, cm^{-2} s^{-1}$.

|  | TESLA | CLIC |
|---|---|---|
| Accelerating gradient MeV/m | 23.4 | 150 |
| Bunch spacing, $\tau_e$ (ns) | 211.37 | 0.66 |
| Number of bunches, $n_b$ | 5600 | 154 |
| Repetition rate, $f_{rep}$, (Hz) | 5 | 200 |
| Number of electrons per bunch, $N_e$ ($10^{10}$) | 2 | 0.4 |

Table II. Nominal parameters of the TESLA and CLIC e-beams

|  | "TESLA" | "CLIC" |
|---|---|---|
| Effective linac length (km) | 2.88 | 0.45 |
| Bunch spacing, $\tau_e$ (ns) | 211.37 | 0.66 |
| $\tau_p$ (ns) | 25 | 25 |
| $\tau_p/\tau_e$ ($\tau_p = 25$ ns) | 0.118 | 37.88 |
| $n_b^{eff}$ | 5600 | 5 |
| $N_p$ | $1.1 \times 10^{11}$ | $1.1 \times 10^{11}$ |
| $\beta_p^*$ (m) | 0.5 | 0.5 |
| $\varepsilon_p$ (nm) | 0.5 | 0.5 |
| $L_{ep}$ ($cm^{-2}s^{-1}$) | $1.9 \cdot 10^{30}$ | $1.4 \cdot 10^{28}$ |

Table III. Main parameters of "TESLA"-LHC and "CLIC"-LHC colliders with nominal LHC beam

|  | "TESLA" | "CLIC" |
|---|---|---|
| $N_p$ | $10^{11}$ ($5 \cdot 10^{11}$) | $10^{11}$ ($5 \cdot 10^{11}$) |
| $\beta_p^*$ (cm) | 10 | 10 |
| $\varepsilon_p^N$ (µm) | 1 | 1 |
| $\Delta Q_p$ | 0.0024 | 0.0005 |
| Disruption D | 12 (60) | 12 (60) |
| $L_{ep}$ ($cm^{-2}s^{-1}$) | $3.3 \cdot 10^{31}$ ($1.6 \cdot 10^{32}$) | $2.4 \cdot 10^{29}$ ($1.2 \cdot 10^{30}$) |

Table IV. Main parameters of "TESLA"-LHC and "CLIC"-LHC with THERA like upgrade of the LHC proton beam parameter

Because of 7 times higher proton beam energy comparing to the HERA the number of protons in LHC bunches can be essentially enlarged. For example, the LHC beam lifetime is ~5 h for $N_p = 5 \times 10^{11}$ and $\varepsilon_p^N = 10^{-6}$ m. Therefore, luminosity $L_{ep} = 10^{32}$ cm$^{-2}$s$^{-1}$ can be achieved with TESLA technology. Radical modification of electron beam is necessary in the case of CLIC technology. For example, $N_e$ can be enlarged by the factor of 2.5 [16] (the beam-beam tune shift, Eq.(3), permits the factor ~6). In addition, the effective collision frequency can be enlarged by factor 10 due to corresponding increase of the number of bunch trains per RF pulse a la CLICHÉ [17]. Remaining factor 4 may be provided by "dynamic focusing" [18].

To summarize, using TESLA and CLIC like electron linacs with active lengths ~2.9 km and ~0.45 km, respectively, one can obtain the same center of mass energy as in the case of ~27 km electron ring. Concerning the luminosity, "moderate" upgrade of TESLA and LHC beams could give opportunity to achieve $L_{ep} = 10^{32}$ cm$^{-2}$s$^{-1}$, whereas "radical" upgrades of e-beam is needed for CLIC.

**4. The VLHC Based ep Collider**

The standard type ep collider based on VLHC assumes a construction of 180 GeV e-ring in the VLHC tunnel [9]. In this case length of the ring is 531 km, radiated power loss for electron beam is 50 MW, center of mass energy is 6 TeV and luminosity is $1.4 \cdot 10^{32}$ cm$^{-2}$s$^{-1}$ [9]. Main parameters of this machine are listed in Table V.

Concerning linac-ring option, with $E_e = 180$ GeV and $P_e = 50$ MW and THERA like upgrade of the VLHC proton beam parameters $N_p = 10^{11}$, $\varepsilon_p^N = 10^{-6}$ m and $\beta_p^* = 10$ cm, according to Eq. (1) we obtain $L_{ep} = 7.3 \cdot 10^{32}$ cm$^{-2}$s$^{-1}$ for "ideal" e-linac. The active lengths are 7.7 km and 1.2 km for TESLA and CLIC like electron linacs, respectively.

| Electron beam parameters | |
|---|---|
| Energy $E_e$ (GeV) | 180 |
| Bunch Population, $N_e$ | $10.1 \times 10^{10}$ |
| Beam-beam tune shifts, $\xi_{e,x}/\xi_{e,y}$ $(10^{-3})$ | 06.1/2.9 |
| Radiation power W [MW] | 50 |
| Proton beam parameter | |
| Energy $E_p$ (TeV) | 50 |
| Bunch Population, $N_p$ | $12.5 \times 10^{10}$ |
| Beam-beam tune shifts, $\xi_{p,x}/\xi_{p,y}$ $(10^{-3})$ | 4/0.3 |
| Number of bunches | 6000 |
| Collider parameters | |
| Center of mass energy, $\sqrt{s}$ (TeV) | 6 |
| Luminosity ($10^{32}$ cm$^{-2}$s$^{-1}$) | 1.4 |
| Circumference, (km) | 531 |

Table V. Parameters for standard type VLHC ep collider [9].

Main parameters of "TESLA"-VLHC and "CLIC"- VLHC options with THERA like upgrade of the VLHC proton beam are given in the Table VI. It is seen that the needed luminosity is achieved with nominal TESLA parameters, whereas a factor of ~70 is required for the CLIC case. Possible solutions for the latter case are presented in the previous section.

| | "TESLA" | "CLIC" |
|---|---|---|
| $N_p$ | $10^{11}$ | $10^{11}$ |
| $\beta_p^*$ (cm) | 10 | 10 |
| $\varepsilon_p^N$ (μm) | 1 | 1 |
| $\Delta Q_p$ | 0.0024 | 0.0005 |
| Disruption D | 31.8 | 31.8 |
| Effective linac length (km) | 7.69 | 1.2 |
| Bunch spacing, $\tau_e$ (ns) | 211.37 | 0.66 |
| $\tau_p$ (ns) | 19 | 19 |
| $\tau_p/\tau_e$ ($\tau_p = 19$ ns) | 0.089 | 28.78 |
| $n_b^{eff}$ | 5600 | 6 |
| $L_{ep}$ (cm$^{-2}$s$^{-1}$) | $2.3 \cdot 10^{32}$ | $2 \cdot 10^{30}$ |

Table VI. Main parameters of "TESLA"-VLHC and "CLIC"-VLHC with THERA like upgrade of the LHC proton beam parameter

## 5. Physics Example: Pair production of heavy quarks

The physics search potential of an $\sqrt{s} = 1\,\text{TeV}$ ep collider is extensively analyzed during 2 years THERA study [13] (see also [19-27]). Here we consider in details one example, namely gluon distributions, which is very important for Higgs physics at the LHC, because gluon fusion is the dominant channel for Higgs boson production at pp colliders.

Measurements of the heavy quarks produced in the process of photon-gluon fusion (PGF) can be used for the direct reconstruction of the gluon structure of the proton [28]. Fig.1 shows the differential cross sections $d\sigma/d\log_{10} x_g$ ($x_g$ denoting the gluon fractional moment in the proton) for charm and beauty produced in PGF at HERA, THERA, QCD Explorer and Linac-VLHC. The cross sections were calculated within NLO QCD [29] for $Q^2 < 1\,\text{GeV}^2$. The GRV98 [30] parameterization was used for the proton structure function. The parameterization was artificially extended to the range $Q^2 > 10^6\,\text{GeV}^2$ to cover the full-scale range of the Linac-VLHC collider. The increase of the electron beam energy will provide an opportunity to probe at THERA one order of magnitude smaller $x_g$ values with respect to those at HERA. The kinematics limits of the $x_g$ measurements at THERA are $10^{-5}$ and $10^{-4}$ for charm and beauty production, respectively. Similar statement is valid for QCD Explorer, which has approximately the same center of mass energy as THERA. Linac-VLHC will give opportunity to explore order lower values of $x_g$. However, to be sensitive to the $x_g$ values around the kinematics limits one will need to tag heavy quarks in the very backward (electron) direction.

Plots (a) and (b) in Fig. 2 show the predictions for THERA with $E_e = 250\,\text{GeV}$ and $E_p = 920\,\text{GeV}$ imposing additional cuts $\theta^{c,b} < 179^0$, $\theta^{c,b} < 175^0$ and $\theta^{c,b} < 170^0$. Only charm quarks with $\theta^c > 175^\circ$ demonstrate sensitivity to the as yet unexplored range $x_g < 10^{-4}$.

Similar plots for QCD Explorer are presented in Fig. 3. Due to a little bit higher center of mass energy and, especially, larger asymmetry of beam energies, one will be able to explore $10^{-5} < x_g < 10^{-4}$ using charm quarks with $\theta^c < 175^0$. Moreover, comparison of Figs. 2(a) and 3(b) show that the $x_g$ region, which can be explored using c quarks with $\theta^c < 175^0$ at THERA, is covered by b quarks with $\theta^b < 175^0$ at QCD Explorer. Approximately ~40 times larger cross-section for $c\bar{c}$ pair production at THERA comparing to $b\bar{b}$ pair production at QCD Explorer can be compensated by higher luminosity of the latter one.

Linac-VLHC will give opportunity to explore $10^{-5} < x_g < 10^{-4}$ using beauty quarks with $\theta^b < 175^0$ as seen from Fig. 4(b), whereas charm quarks are sensitive up to an order lower $x_g \approx 10^{-6}$ (Fig 4(a)).

**6. Conclusion**

Lepton hadron colliders with $\sqrt{s} < 1$ TeV are necessary both to clarify fundamental aspects of the QCD part of the Standard Model and for adequate interpretation of experimental data from multi-TeV hadron colliders. A construction of an additional e-ring in the LHC and VLHC tunnels might cause a lot of technical problems (an example is inevitable removing of the LEP from the tunnel in order to assemble the LHC). Linacs give opportunity to obtain the same e-beam energy with much shorter lengths.

In spite of approximately equal center of mass energies, QCD Explorer is more advantageous then THERA for exploration of small $x_g$ region. It is important to compare the physics search potential of these machines for other topics. Obviously linac-VLHC has larger physics search potential.

Finally, electron beam energy can be expanded by increasing linac lengths, whereas synchrotron radiation blocks this road for standard type ep colliders.

**Acknowledgments:** This work is partially supported by Turkish State Planning Organization under the grand no 2002K120250.

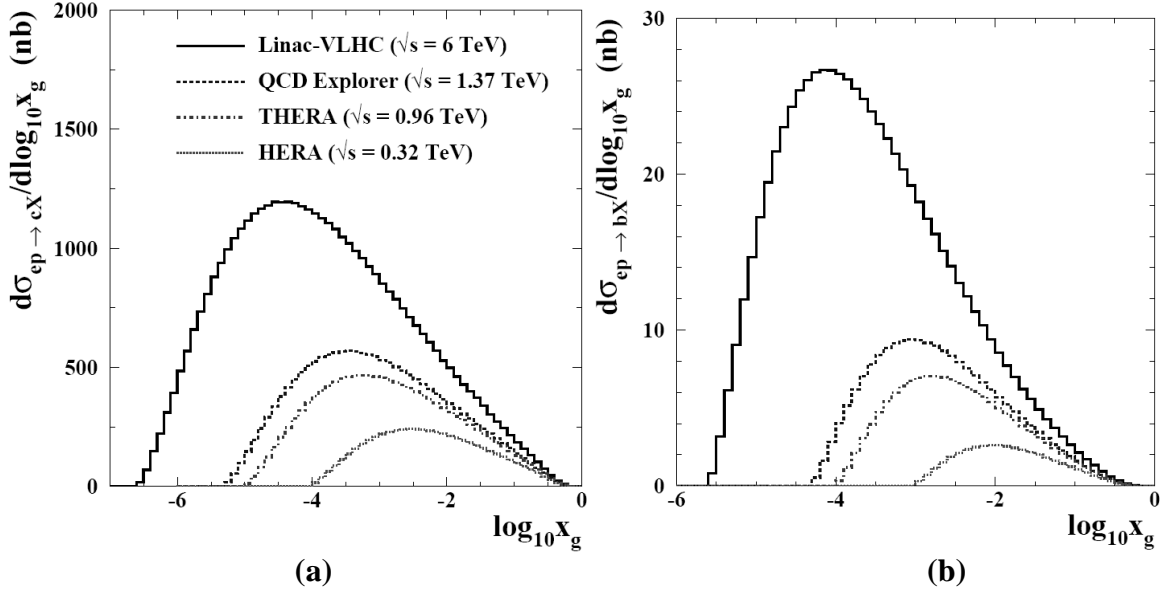

**Figure 1.** The differential cross-section $d\sigma/d\log_{10} x_g$ for charm (a) and beauty (b) produced in the process of $\gamma^* g$ fusion. Solid, dash-dotted, dashed and dotted curves correspond to Linac-VLHC, QCD Explorer, THERA and HERA, respectively.

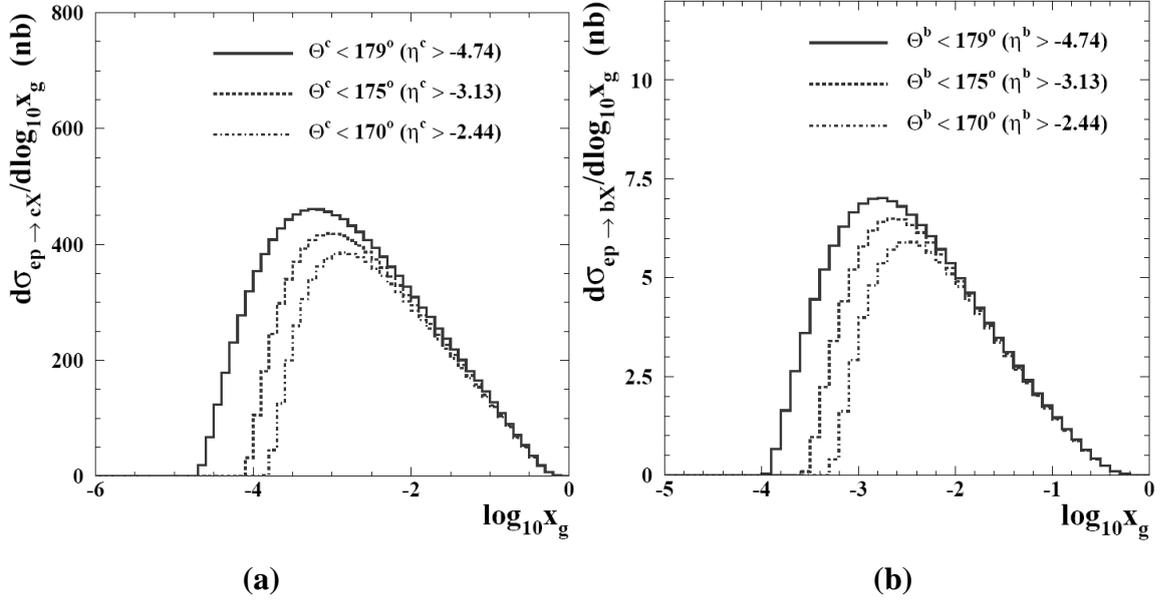

**Figure 2.** The prediction for the THERA with additional cut $\theta^{c,b} < 179^0$ (solid curves), $\theta^{c,b} < 175^0$ (dashed curves) and $\theta^{c,b} < 170^0$ (dotted curves)

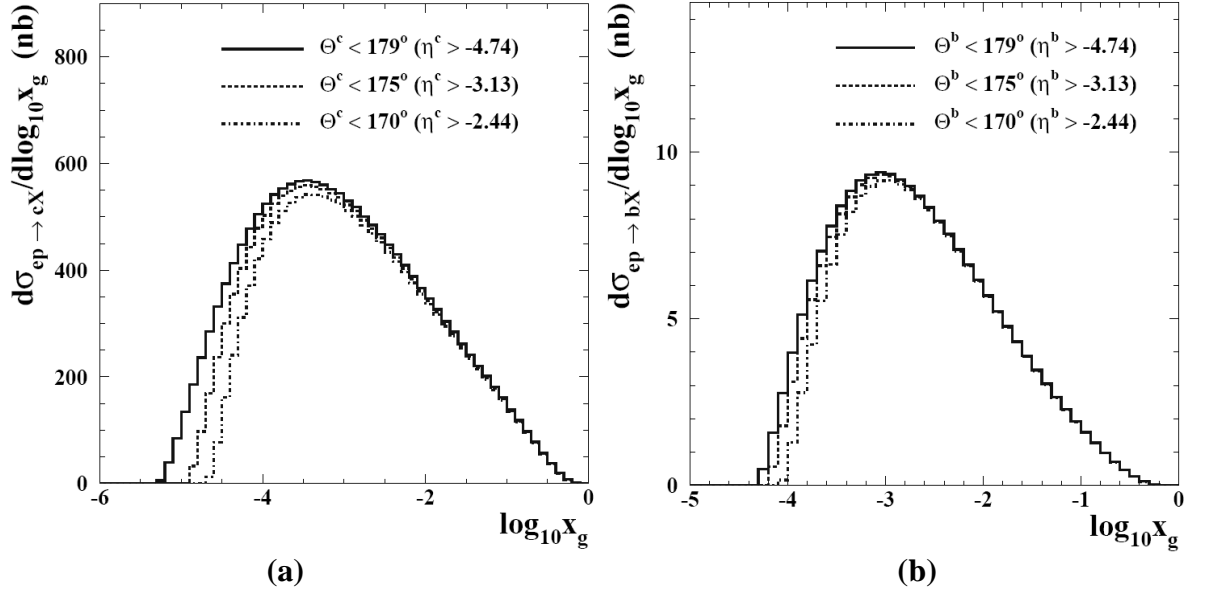

**Figure 3.** The same as Fig.2 but for the QCD Explorer.

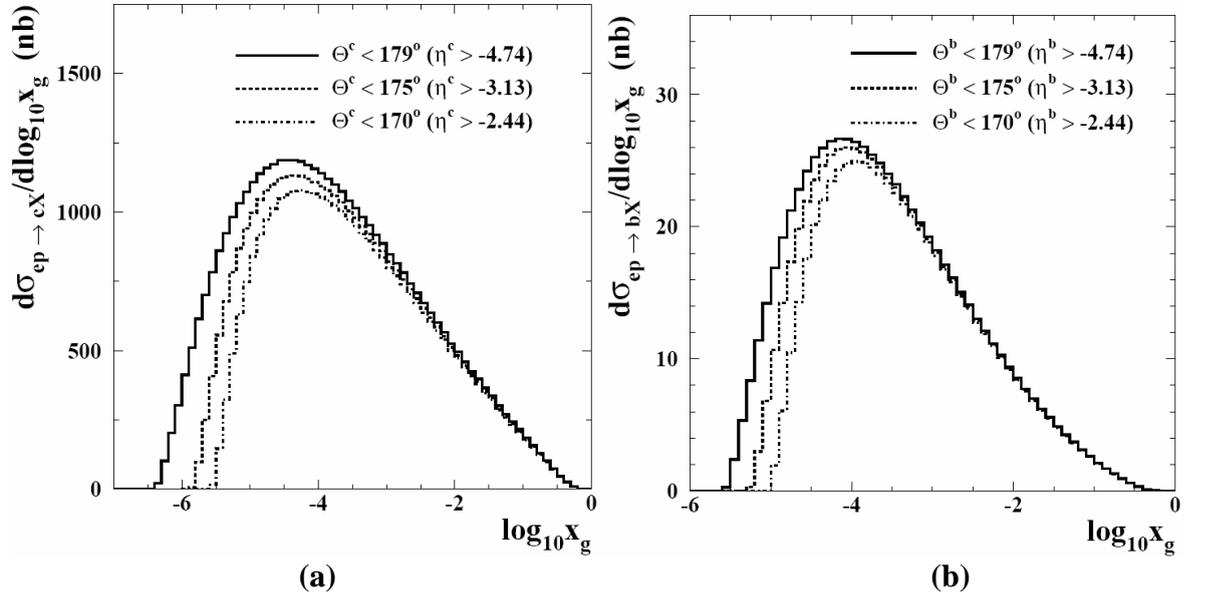

**Figure 4.** The same as Fig. 2 but for Linac-VLHC.